\documentclass[letterpaper, 10 pt, conference]{IEEEtran}
\IEEEoverridecommandlockouts
 \usepackage{booktabs} 
\usepackage{cite}
\usepackage{algorithmic}
\usepackage[algoruled]{algorithm2e}
\usepackage{textcomp}
\usepackage{xcolor}
\usepackage{bm}
\usepackage{amsfonts}
\usepackage{amsmath,cases,amssymb}
\usepackage[english]{babel}
\usepackage{cite}
\usepackage{color}
\usepackage{nomencl}
\usepackage{url}
\usepackage{bm}
\usepackage{multirow}
\usepackage{hyperref}
\usepackage{color}
\usepackage{multirow}
\usepackage{mathtools}
\usepackage{etoolbox}
\newtheorem{myth}{Theorem}

\DeclareMathSymbol{\Minus}{\mathbin}{AMSa}{"39}

\def\Plus{\texttt{+}}
\begin{document}

\title{\LARGE \bf
Prescribing Decision Conservativeness in Two-Stage Power Markets: A Distributionally Robust End-to-End Approach
}

\author{Zhirui Liang$^{1}$, Qi Li$^{2}$,  Anqi Liu$^{3}$, Yury Dvorkin$^{1,4}$ \\
\textit{$^1$ Department of Electrical and Computer Engineering, Johns Hopkins University}\\
\textit{$^2$ Department of Applied Mathematics and Statistics, Johns Hopkins University}\\
\textit{$^3$ Department of Computer Science, Johns Hopkins University}\\
\textit{$^4$ Department of Civil and System Engineering, Johns Hopkins University}\\
Baltimore MD, U.S. \\
zliang31@jhu.edu, qli112@jhu.edu, aliu@cs.jhu.edu, ydvorki1@jhu.edu
\thanks{Zhirui Liang and Qi Li contributed equally to this work. Corresponding author: Zhirui Liang}%
}

\maketitle

\begin{abstract}
This paper presents an end-to-end framework for calibrating wind power forecast models to minimize operational costs in two-stage power markets, where the first stage involves a distributionally robust optimal power flow (DR-OPF) model. Unlike traditional methods that adjust forecast parameters and uncertainty quantification (UQ) separately, this framework jointly optimizes both the forecast model parameters and the decision conservativeness, which determines the size of the ambiguity set in the DR-OPF model. The framework aligns UQ with actual uncertainty realizations by directly optimizing downstream operational costs, a process referred to as cost-oriented calibration. The calibration is achieved using a gradient descent approach. To enable efficient differentiation, the DR-OPF problem is reformulated into a convex form, and the Envelope Theorem is leveraged to simplify gradient derivation in the two-stage setting. Additionally, the framework supports distributed implementation, enhancing data privacy and reducing computational overhead. By proactively calibrating forecast parameters and prescribing optimal decision conservativeness, the framework significantly enhances cost efficiency and reliability in power system operations. Numerical experiments on an IEEE 5-bus system demonstrate the effectiveness and efficiency of the proposed approach.
\end{abstract}

\section{Introduction}
Decision-focused learning (DFL) is an emerging approach that combines machine learning (ML) with constrained optimization to minimize decision regret under uncertainty \cite{mandi2024decision}. DFL replaces or augments the standard ML training loss with task-specific loss and is trained end-to-end, leveraging feedback from downstream optimization through modified gradient descent. In power systems, DFL is applied to generate extreme uncertainty scenarios that challenge operations \cite{liang2022operation}, design false data injection attacks that increase operational costs \cite{chen2022vulnerability}, and, more commonly, make predictions to minimize downstream costs \cite{donti2017task,zhang2024toward,dvorkin2024regression}.

This paper examines the forecast-then-optimize framework, where renewable or load forecasts are generated and then used in downstream power system operations. The operation problem is typically formulated as a two-stage optimization, with schedules set during the look-ahead stage and adjusted in real-time dispatch. Due to the asymmetric impact of over-prediction and under-prediction on total operational costs, more accurate forecasts do not always result in better decisions. Previous studies \cite{morales2023prescribing,zhang2024toward,dvorkin2024regression} have focused on adjusting the forecast parameter $\theta$ to minimize total costs across two stages but have overlooked the role of uncertainty quantification (UQ), which critically impacts costs by determining the flexibility reserves scheduled in the first stage. While some works \cite{mieth2024prescribed,wang2024learning} have explored prescribing the size of uncertainty sets within a DFL framework, they did not address adjustments to the forecast model itself. Since the total cost in a two-stage problem depends on both the point forecast and UQ, we argue that co-optimizing parameters for both is a more effective approach.

Distributionally robust optimization (DRO), which identifies solutions that perform well across multiple distributions within ambiguity sets, is widely used in decision-making under uncertainty. In power system operations, distributionally robust optimal power flow (DR-OPF) can replace deterministic reserve scheduling models in the look-ahead stage to better manage uncertainty \cite{mieth2023data}. The size of the ambiguity set in a DRO model, denoted as $\epsilon$, reflects the \textit{decision conservativeness} of the decision maker, with larger values of $\epsilon$ indicating a more conservative approach \cite{li2024revealing,liang2024learning}.
In the forecast-then-optimize framework, system operators may require forecasting agents to submit the value of $\epsilon$ along with their forecasts, where $\epsilon$ represents the \textit{data quality} of the forecast \cite{mieth2023data}. However, agents may not report their true data quality, and even if they do, the reported $\epsilon$ may not be optimal for downstream tasks since agents lack access to these tasks and their incentives are not aligned with them. To address this, we propose a distributionally robust end-to-end framework that enables system operators to prescribe their decision conservativeness in a two-stage power market. 

Several studies have explored the DFL framework within a DRO context. For instance, \cite{costa2023distributionally} co-optimize prediction parameters along with those controlling risk appetite and model robustness in a $\phi$-divergence DRO, while \cite{ma2024differentiable} formulate generic mixed-integer DRO as a differentiable layer. Unlike these works, where DRO is the sole downstream task, our problem positions DRO as the first stage in a two-stage optimization framework. 
We first reformulate the Wasserstein-distance-based DRO into a convex form \cite{mohajerin2018data}, aligning the task of differentiating through DRO with that of differentiating through generic convex problems \cite{amos2017optnet,agrawal2019differentiable}. Additionally, we leverge the Envelope Theorem to simplify gradient derivation in two-stage problems \cite{takayama1985mathematical}.

In summary, this paper proposes an end-to-end framework for power system operators to calibrate wind power forecast models by minimizing the downstream operational cost in a two-stage market. This framework addresses two critical aspects: adjusting the parameter $\theta$ of a pretrained forecast model and determining the decision conservativeness $\epsilon$. The objective is not merely to improve the accuracy of UQ but to ensure it aligns with the operational needs of the system by reflecting actual uncertainty realizations. Unlike traditional calibration methods that minimize prediction errors, this approach quantifies the alignment between forecasted and realized uncertainty based on their impact on operational costs, a process termed \textit{cost-oriented calibration}.

Unlike our previous work \cite{liang2024learning}, which adjusts the value of $\epsilon$ retrospectively after decisions are made and uncertainty is realized, this paper focuses on determining the optimal value of $\epsilon$ proactively, before operational decisions are made. This approach accounts for potential uncertainty realizations in advance, making the process inherently \textit{prescriptive}.
The proposed framework can be implemented in a distributed manner, which enhances information privacy of forecasting agents and reduces the system operator's computational burden.

\section{Operations of Two-Stage Power Markets}
\subsection{Wind Power Forecasting and Uncertainty Quantification} \label{sec:: prediction model}
Consider a power system with $N_w$ wind farms, where each farm independently serves as a forecasting agent to predict its own wind power output.
The $j^\text{th}$ forecasting agent trains a machine learning model $\hat y_j= W_{\theta_j}(x_j)$, where $\hat y_j$ is the forecasted wind power output, $x_j$ collects relevant features (e.g., meteorological data), and $\theta_j$ is the model parameter. Following \cite{dvorkin2024regression}, we adopt a regression model $\hat y_j = \theta_j^{\top} x_j$ for simplicity, though more complex models like neural networks can also be applied. Forecast models are typically trained by minimizing forecast errors, such as the mean squared error (MSE).

Let $\xi_j$ denote the stochastic forecast error of the  $j^\text{th}$ forecast model, which can be characterized by its empirical distribution. Using a dataset of $N_s$ historical samples $\left\{(x_{ji}, y_{ji})\right\}_{i=1}^{N_s}$, we compute a set of historical forecast errors $\{\hat{\xi}_{ji}\}_{i=1}^{N_s}$, where $\hat{\xi}_{ji} = y_{ji} - \theta_j^{\top} x_j$. These forecast errors define the empirical distribution of $\xi_j$ as:
\begin{equation}
    \label{def:: empirical distribution}
    \hat{\mathbb{P}}_{j} (\theta_j) = \frac{1}{N_s}\sum\nolimits_{i=1}^{N_s} \delta_{\hat \xi_{ji}},
\end{equation}
where $\delta_{\hat{\xi}_{ji}}$ is a Dirac distribution centered at $\hat{\xi}_{ji}$. 

Given the limited availability of historical data, the empirical distribution $\hat{\mathbb{P}}_{j} (\theta_j)$ may not fully capture the true distribution of $\xi_j$. 
Nevertheless, the true distribution of $\xi_j$, denoted as $\mathbb{P}_j(\theta_j)$, should lie within a bounded distance from $\hat{\mathbb{P}}_{j} (\theta_j)$. Assuming this bounded distance is defined using the Wasserstein metric \cite{mohajerin2018data}, the true distribution is assumed to belong to the ambiguity set $\mathcal{B}_{\epsilon_j}\left(\hat{\mathbb{P}}_{j} (\theta_j)\right)$, given by:
\begin{equation} \label{def:: Ambiguity set}
    \mathcal{B}_{\epsilon_j}\left(\hat{\mathbb{P}}_{j} (\theta_j)\right) = \left\{\mathbb{Q} \in \mathcal{M}(\Xi_j) \Big\vert d_{W}(\hat{\mathbb{P}}_{j} (\theta_j), \mathbb{Q}) \leq \epsilon_j \right\},
\end{equation}
where $\Xi_j$ is the domain of the random variable $\xi_j$, $\mathcal{M}(\Xi_j)$ collects all distributions on $\Xi_j$, $d_{W}(\mathbb{P}_1, \mathbb{P}_2): \mathcal{M}(\Xi_j) \times \mathcal{M}(\Xi_j) \rightarrow \mathbb{R}$ measures the Wasserstein distance between two distributions $\mathbb{P}_1$ and $\mathbb{P}_2$, $\epsilon_j$ is the Wasserstein ball radius which controls the size of the ambiguity set. A larger $\epsilon_j$ indicates a greater deviation of the true distribution $\mathbb{P}_j$ from the empirical distribution $\hat{\mathbb{P}}_{j}$, implying less consistent performance of the forecast model. 



\subsection{Reserve Procurement in Look-Ahead Scheduling} \label{sec:: day-ahead market}
Forecasted wind power is used in look-ahead markets (e.g., day-ahead, intra-day, hour-ahead) to schedule power generation and reserves from conventional generators, forming the first stage of a two-stage power market. In a transmission network with $N_b$ buses, $N_g$ generators, $N_w$ wind farms, $N_l$ transmission lines, and demand $d \in \mathbb{R}^{N_b}$, forecasted wind power $\hat{y} \in \mathbb{R}^{N_w}$ may contain forecast error $\xi$, yielding actual wind power $\hat{y} + \xi$. To maintain power balance, generator outputs are given by $g - A\xi$, where $g \in \mathbb{R}^{N_g}$ is the scheduled generation, and $A \in \mathbb{R}^{N_g \times N_w}$ represents reserve participation factors for managing wind power deviations.
Following \cite{mieth2023data}, system operators can construct the \textit{multi-source Wasserstein ambiguity set} based on the forecast from all wind farms:
\begin{equation} \label{def:: MSW Ambiguity set}
\mathcal{A}^{\text{MSW}} := \left\{ \mathbb{Q} \in \mathcal{M}(\Xi) \;\middle|\; 
\begin{aligned}
    &\mathbb{P}_{j\#} \mathbb{Q} = \mathbb{Q}_j,\ j=1,...,N_w \\
    &d_{W}(\hat{\mathbb{P}}_{j}, \mathbb{Q}_j) \leq \epsilon_j,\ j=1,...,N_w
\end{aligned}
\right\}
\end{equation}
where $\xi=[\xi_1,...,\xi_{N_w}]$ collects the forecast error of all the wind farms, $\Xi$ is the domain of $\xi$, $\mathbb{P}_{j\#}$ denotes the push-forward distribution of the joint measure $\mathbb{Q}$ under the projection onto the $j^\text{th}$ coordinate. Based on this ambiguity set, the \textit{distributionally robust optimal power flow} (DR-OPF) model is formulated as:
\begin{subequations}
    \label{DA:: total original}
    \begin{align}
        \min_{g, r^\Plus, r^\Minus, A} \quad & c_{g}^{\top} g + c_{r}^{\top}(r^\Plus + r^\Minus) + \sup_{\mathbb{Q} \in \mathcal{A}^{\text{MSW}}} \mathbb{E}^{\mathbb{Q}}\left[c_{a}^{\top} (\Minus A \xi) \right] \label{DA:: objective}\\
        \mathrm{s.t.} \quad & \mathbf{1}_{g}^{\top}g + \mathbf{1}_{w}^{\top}\hat y = \mathbf{1}_{b}^{\top}d \label{DA:: power balance} \\
                            & \underline{g} + r^\Minus \leq g \leq \overline{g} - r^\Plus\label{DA:: gen bounds}\\
                            & \Minus \overline{f} \leq \Phi\left[S_g g + S_w \hat y -d\right] \leq \overline{f} \label{DA:: power flow constraints} \\
                            & A^{\top} \mathbf{1}_{g}= \mathbf{1}_{w} \label{DA:: distribution factor sum}\\
                            & \inf_{\mathbb{Q} \in \mathcal{A}^{\text{MSW}}}\text{Pr}^{\mathbb{Q}} \left\{ -r^\Minus \leq A \xi \leq r^\Plus \right\} \geq 1 - \gamma \label{DA:: chance constraint}\\
                            & r^\Plus, r^\Minus, A \geq 0, \label{DA:: reserve positive} 
    \end{align}
\end{subequations}
where $c_g, c_r$, and $c_a$ are the unit costs of scheduled power generation, scheduled reserve, and reserve activation, respectively. 
Parameters $\underline{g}, \overline{g} \in \mathbb{R}^{N_g}$ define the generation bounds, $\overline{f} \in \mathbb{R}^{N_l}$ is the power flow limit, and $\Phi \in \mathbb{R}^{N_b \times N_l}$ is the power transfer distribution factor (PTDF) matrix. Indicator matrices $S_g$ and $S_w$ link generators and wind farms to buses, while $\mathbf{1}_g$, $\mathbf{1}_w$, and $\mathbf{1}_b$ are all-one vectors of dimension $N_g$, $N_w$, and $N_b$. Decision variables include $g$, $A$, and $r^\Plus, r^\Minus \in \mathbb{R}^{N_g}$, representing scheduled generation, reserve participation factors, and upward/downward reserves, respectively.

The DR-OPF model in \eqref{DA:: total original} determines the least-cost schedule under the worst-case scenario, where the realization of $\xi$ incurs the highest real-time cost. Objective \eqref{DA:: objective} minimizes total costs, including generation, reserve procurement, and worst-case expected reserve activation. The first two terms represent actual look-ahead market costs, while the third reflects anticipated real-time dispatch costs. Constraint \eqref{DA:: power balance} ensures power balance, \eqref{DA:: gen bounds} enforces generator operational limits, and \eqref{DA:: power flow constraints} ensures compliance with thermal limits. Constraint \eqref{DA:: distribution factor sum} aligns reserve deployment with expected wind power fluctuations. The chance constraint in \eqref{DA:: chance constraint} limits the violation rate of reserve scheduling to a predefined risk level $\gamma$.

The DR-OPF model in \eqref{DA:: total original} can be reformulated into a tractable convex optimization problem, as outlined in \cite{mohajerin2018data}. For brevity, we omit the detailed derivation steps and present the reformulated problem directly as follows:
\begin{subequations}
\allowdisplaybreaks
    \label{RDA:: total}
    \begin{align} 
        \min_{G, A, \Lambda} \  &f_1 = c_{g}^{\top} g + c_{r}^{\top}(r^\Plus + r^\Minus) + \sum_{j=1}^{N_w} \left(\lambda_j^{O} \epsilon_j + \frac{1}{N_s}\sum_{i=1}^{N_s} s^{O}_{ji} \right) \label{RDA:: objective}\\
        \mathrm{s.t.} \  & \eqref{DA:: power balance} - \eqref{DA:: distribution factor sum}  \nonumber\\
        & s_{ji}^{O} \geq a_{j}^{O} \hat \xi_{ji} \label{RDA:: auxiliary obj 1}\\
        & s_{ji}^{O} \geq a_{j}^{O} \overline{\xi}_j - \lambda_j^{O}(\overline{\xi}_j - \hat \xi_{ji}) \label{RDA:: auxiliary obj 2}\\
        & s_{ji}^{O} \geq a_{j}^{O}\underline{\xi}_j + \lambda_j^{O}(\underline{\xi}_j - \hat \xi_{ji}) \label{RDA:: auxiliary obj 3}\\
        & \tau + \frac{1}{\gamma} \left(\sum\nolimits_{j=1}^{N_w} \lambda_j^{C} \epsilon_j + \frac{1}{N_s} \sum\nolimits_{i=1}^{N_s} s^{C}_{i} \right)\leq 0 \label{RDA:: reformulated chance constraints}\\
        & s_{i}^{C} \geq b_k^{C} + \sum\nolimits_{j=1}^{N_w} s_{jik}^{C} \label{RDA:: auxiliary CC 1}\\
        & s_{jik}^{C} \geq a_{kj}^{C} \hat \xi_{ji} \label{RDA:: auxiliary CC 2}\\
        & s_{jik}^{C} \geq a_{kj}^{C} \overline{\xi}_j - \lambda_j^{C} (\overline{\xi}_j - \hat \xi_{ji}) \label{RDA:: auxiliary CC 3}\\
        & s_{jik}^{C} \geq a_{kj}^{C} \underline{\xi}_j + \lambda_j^{C} (\underline{\xi}_j - \hat \xi_{ji}) \label{RDA:: auxiliary CC 4}\\
        & r^\Plus, r^\Minus, A, \lambda_j^{O}, \lambda_j^{C} \geq 0 \label{RDA:: auxiliary constraints end}\\
        & \forall i = 1,...,N_s, \ j = 1, ..., N_w, \ k = 1,..., 2N_g+1, \nonumber  
    \end{align}
\end{subequations}
where $G$ contains the primal variables $g$, $r^\Plus$, and $r^\Minus$, while $\Lambda$ includes auxiliary variables $\tau$, $\lambda^{O}$, $\lambda^{C}$, $s^{O}$, and $s^{C}$, which are introduced during the reformulation of \eqref{DA:: total original}. The superscripts in $\lambda^{O}$ and $s^{O}$ relate to the objective function reformulation, and those in $\lambda^{C}$ and $s^{C}$ relate to the chance-constraint reformulation. Eqs.\eqref{RDA:: objective}-\eqref{RDA:: auxiliary obj 3} provide an \textit{exact} reformulation of \eqref{DA:: objective}, while Eqs.\eqref{RDA:: reformulated chance constraints}-\eqref{RDA:: auxiliary CC 4} offer an \textit{inner} approximation of \eqref{DA:: chance constraint}. Bounds $\underline{\xi}$ and $\overline{\xi}$ on $\xi$ are inferred from datasets, and $a_k^C$, $b_k^C$, and $a_k^O$ correspond to the $k$-th row or entry of the following matrix and vector:
\begin{equation}
    \nonumber
    \label{Notation:: RDA}
    \left[
        \begin{array}{c}
             0 \\
             A \\
             \Minus A
        \end{array}
    \right], 
    \left[
        \begin{array}{c}
             0 \\
             \Minus r^\Plus - \tau \mathbf{1}_{N_g} \\
             \Minus r^\Minus - \tau \mathbf{1}_{N_g}
        \end{array}
    \right], \text{and} \Minus A^{T}c_a.
\end{equation}

\subsection{Reserve Activation in Real-Time Dispatch} \label{sec:: real-time market}
The second stage of a two-stage power market adjusts flexible resource dispatch to maintain real-time power balance. When actual wind generation $y$ deviates from the forecast $\hat y$, controllable generators will provide upward or downward flexibility. Denote the optimal schedule from \eqref{RDA:: total} as $g^*$, $r^{\Plus*}$, and $r^{\Minus*}$. If real-time power output of generators fall within $g^* - r^{\Minus*}$ and $g^* + r^{\Plus*}$, scheduled flexibility is sufficient; otherwise, costly emergency responses are required. The real-time dispatch aims to minimize adjustment costs while ensuring power balance, which can be formulated as:
\begin{subequations}
    \allowdisplaybreaks
    \label{RT:: total}
    \begin{align}
     &\min_{r_{\text{in}}, r_{\text{out}}^{\Plus}, r_{\text{out}}^{\Minus}} \quad f_2= c_{\text{in}}^{\top} \vert r_{\text{in}} \vert + c_{\text{out}}^{\Plus \top} r^\Plus_{\text{out}} + c_{\text{out}}^{\Minus \top} r^\Minus_{\text{out}} \label{RT:: obj}\\
    & \mathrm{s.t.} \ (\lambda): \mathbf{1}_{g}^{\top} g^{*} + \mathbf{1}^{\top}_{g} ( r_{\text{in}} +  r_{\text{out}}^{\Plus} -  r_{\text{out}}^{\Minus} ) + \mathbf{1}_{w}^{\top} y = \mathbf{1}_{l}^{\top} d \label{RT:: power balance}\\
    & (\underline{\mu}, \overline{\mu}): \underline{g} - g^{*} \leq r_{\text{in}} + r_{\text{out}}^{\Plus} - r_{\text{out}}^{\Minus} \leq \overline{g} - g^{*} \label{RT:: generator bounds}\\
    & (\underline{\phi}, \overline{\phi}): \left\vert \Phi \left[S_g(g^* + r_{\text{in}} + r_{\text{out}}^{\Plus} - r_{\text{out}}^{\Minus}) + S_w y - d \right]\right\vert \leq \overline{f} \label{RT:: power flow} \\
    &(\underline{\nu}, \overline{\nu}): \Minus r^{\Minus*} \leq r_{\text{in}} \leq r^{\Plus*}  \label{RT:: inner reserve}\\
    & \quad \qquad r_{\text{out}}^{\Plus}, r_{\text{out}}^{\Minus} \geq 0, \label{RT:: outer reserve}
    \end{align}
\end{subequations}
where $r_{\text{in}}$, $r_{\text{out}}^{\Plus}$, and $r_{\text{out}}^{\Minus}$ represent power within the scheduled reserve and excess power beyond the upper and lower reserves, respectively, with costs $c_{\text{in}}$, $c_{\text{out}}^{+}$, and $c_{\text{out}}^{-}$. Excess power costs are normally higher, i.e., $c_{\text{in}} < c_{\text{out}}^{+}, c_{\text{out}}^{-}$. The objective in \eqref{RT:: obj} minimizes total dispatch costs. Constraints \eqref{RT:: power balance}-\eqref{RT:: outer reserve} ensure power balance, enforce generator bounds, maintain power flow limits, and define feasible regions for $r_{\text{in}}$, $r_{\text{out}}^{\Plus}$, and $r_{\text{out}}^{\Minus}$. Greek letters on the left side of \eqref{RT:: power balance}-\eqref{RT:: inner reserve} denote dual variables for the respective constraints.

The link between look-ahead scheduling and real-time dispatch introduces a trade-off between costs at the two stages. In the DR-OPF model, considering a wider distribution of $\xi$ (i.e., a larger decision conservativeness $\epsilon = [\epsilon_1, \dots, \epsilon_{N_w}]$) results in scheduling more reserve $r^{\Plus*}$ and $r^{\Minus*}$ at the look-ahead stage, increasing scheduling costs but potentially reducing real-time dispatch costs. Additionally, the wind power forecast $\hat{y}$ determines $g^*$, influencing costs at both stages. Consequently, both $\epsilon$ and the forecast model parameter $\Theta = \text{stack}(\theta_1, \dots, \theta_{N_w})$ can be tuned to reduce costs, where $\text{stack}(v_1, v_2)$ vertically combines column vectors $v_1$ and $v_2$. The next section introduces an end-to-end framework for jointly optimizing $\epsilon$ and $\Theta$ to minimize total costs across both stages.

\section{Cost-Oriented Forecast Model Calibration}

\subsection{Overview of the Proposed Framework} 
\label{sec:: DFL Framework}
In power systems, wind power forecasting is typically conducted locally by specialized forecasting agents, with system operators generally not involved in training the forecast model. Assume the agents have pretrained the model $\hat y= \Theta x$ using $N_t$ training data pairs to minimize forecast error. Now, consider a calibration dataset $\mathcal{D}_{c}$ containing $N_c$ data pairs $(x_{i},y_{i})$, where $i \in \mathcal{I}_{c}$. Typically, $N_t \gg N_c$. System operators can perform two tasks using $\mathcal{D}_{c}$:
\begin{itemize}
    \item \textbf{Forecast Model Adaptation}: Adjust $\Theta$ to balance forecast accuracy with the total cost across both stages.
    \item \textbf{Decision Conservativeness Prescription}: Determine the value of $\epsilon$ that minimizes the total cost across both stages.
\end{itemize}

\begin{figure}[!htbp]
    \centering
    \includegraphics[width=1\linewidth]{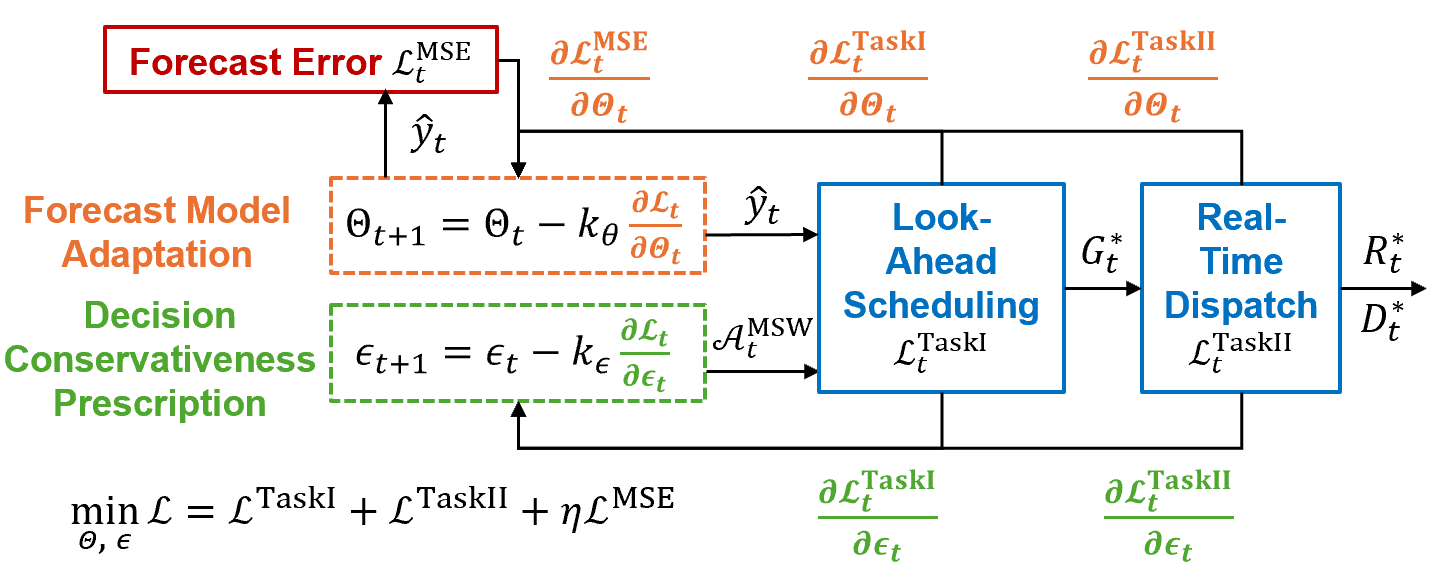}
    \caption{End-to-End Forecast Model Calibration Framework.}
    \label{fig:ETE_framework}
\end{figure}

These two tasks can be formulated as a co-optimization problem involving $\Theta$ and $\epsilon$. The objective is to minimize a composite loss function $\mathcal{L}$, which combines the forecast error $\mathcal{L}^{\text{MSE}}$, the task loss $\mathcal{L}^{\text{TaskI}}$ from look-ahead scheduling, and the task loss $\mathcal{L}^{\text{TaskII}}$ from real-time dispatch. Fig.~\ref{fig:ETE_framework} illustrates an iterative gradient-based approach to achieve these tasks, where where $t$ is the index of iteration, $k_\theta$ and $k_\epsilon$ are the learning rates for $\Theta$ and $\epsilon$, respectively. Moreover, $G^*=stack (g^*, r^{\Plus*}, r^{\Minus*})$ collects variables from the optimal solution of \eqref{RDA:: total}, $R^* = stack (r_{\text{in}}^*, r_{\text{out}}^{\Plus*}, r_{\text{out}}^{\Minus*})$ is the optimal primal solution of \eqref{RT:: total}, and $D^* =stack (\lambda^*, \underline{\mu}^*, \overline{\mu}^*, \underline{\phi}^*, \overline{\phi}^*, \underline{\nu}^*, \overline{\nu}^*)$ is the optimal dual solution of \eqref{RT:: total}. 

In the following sections, we construct the loss functions in Section~\ref{sec:: Loss Function} and derive the corresponding gradients in Section~\ref{sec:: Differentiable Optimization}. To align with the actual data flow in power systems, we further extend the proposed end-to-end framework to a distributed computation approach in Section~\ref{sec:: Distributed Model Calibration}. For simplicity, we aggregate the values from all wind farms and remove the wind farm index $j$ from the notations.

\subsection{Objective Function Construction} 
\label{sec:: Loss Function}
Following \cite{dvorkin2024regression}, we quantify the forecast error using MSE:
\begin{equation} \label{def:: prediction loss}
    \mathcal{L}^{\text{MSE}} = \frac{1}{N_c} \sum\nolimits_{i \in \mathcal{I}_{c}} \left\Vert y_{i} - \Theta^{\top} x_i \right\Vert_{2}^{2}.
\end{equation}

The first task loss $\mathcal{L}^{\text{TaskI}}$, based on cost at the look-ahead scheduling stage, is defined as:
\begin{equation}
    \label{def:: DA task loss}
    \mathcal{L}^{\text{TaskI}} = \frac{1}{N_c}\sum\nolimits_{i \in \mathcal{I}_{c}} c_{g}^{\top} g_i^* + c_{r}^{\top}(r_i^{\Plus*} + r_i^{\Minus*}).
\end{equation}
Note that $\mathcal{L}^{\text{TaskI}}$ includes only the first two terms from the objective function \eqref{RDA:: objective}, representing the actual costs incurred in the look-ahead market.

The second task loss, $\mathcal{L}^{\text{TaskII}}$, represents the reserve activation cost at the real-time stage and is defined as:
\begin{equation}
    \label{def:: real-time task loss}
    \mathcal{L}^{\text{TaskII}} = \frac{1}{N_c}\sum\nolimits_{i \in \mathcal{I}_{c}} c_{\text{in}}^{\top} r_{\text{in},i}^* + c_{\text{out}}^{\Plus \top} r^{\Plus*}_{\text{out},i} + c_{\text{out}}^{\Minus \top} r^{\Minus*}_{\text{out},i}.
\end{equation}

By combining the three loss functions with a weight $\eta$, we formulate the following optimization problem:
\begin{subequations}
\allowdisplaybreaks
    \label{Tri-level problem}
    \begin{align}
        \text{UL}: \min_{\Theta, \epsilon} \ & \mathcal{L}= \mathcal{L}^{\text{TaskI}}(\Theta, \epsilon){\rm{+}} \mathcal{L}^{\text{TaskII}}(\Theta, \epsilon) {\rm{+}} \eta \mathcal{L}^{\text{MSE}}(\Theta) \label{Tri-level:upper}\\
        \mathrm{s.t.} \ & \forall (x_{i},y_{i}) \in \mathcal{D}_{c}, \ i \in \mathcal{I}_{c} \nonumber \\
        & \text{ML}: R_{i}^{*},D_{i}^{*}= \arg \min f_2 \left(G_i^{*}, y_i \right)\label{Tri-level:middle}\\
        & \text{LL}: G_{i}^{*}= \arg \min f_1 \left(\hat y_i, \mathcal{A}^{\text{MSW}} ( \epsilon) \right) \label{Tri-level:lower} \\
        & \hat y_i = \Theta^{\top} x_i. \label{Tri-level:forecast} 
    \end{align}
\end{subequations}
Model \eqref{Tri-level problem} is essentially a tri-level optimization problem. The lower-level (LL) problem  \eqref{Tri-level:lower} is the DR-OPF problem, which depends on $\hat y$ and the ambiguity set $\mathcal{A}^{\text{MSW}}$ constructed based on $\epsilon$. The middle-level (ML) problem \eqref{Tri-level:middle} is real-time dispatch, which requires the optimal schedule $G^*$ from the lower-level problem. Finally, the upper-level (UL) problem \eqref{Tri-level:upper} calibrates the forecast model parameter $\Theta$ and determines the decision conservativeness $\epsilon$.

\subsection{Gradient Descent Approach} \label{sec:: Differentiable Optimization}
Due to numerical challenges associated with directly solving the tri-level problem \eqref{Tri-level problem}, we adopt a gradient-based approach to iteratively find the optimal values of $\Theta$ and $\epsilon$ iteratively. The gradients of $\mathcal{L}$ with respect to $\Theta$ and $\epsilon$ can be computed as:
\begin{subequations}
    \label{def:: total loss gradient}
    \begin{align}
        \frac{\partial \mathcal{L}(\Theta, \epsilon)}{\partial \Theta} &= \frac{\partial \mathcal{L}^{\text{TaskI}}}{\partial \Theta} + \frac{\partial \mathcal{L}^{\text{TaskII}}}{\partial \Theta} + \eta \frac{\partial \mathcal{L}^{\text{MSE}}}{\partial \Theta} \label{def:: total gradient theta}\\
        \frac{\partial \mathcal{L}(\Theta, \epsilon)}{\partial \epsilon} &= \frac{\partial \mathcal{L}^{\text{TaskI}}}{\partial \epsilon} + \frac{\partial \mathcal{L}^{\text{TaskII}}}{\partial \epsilon}, \label{def:: total gradient epsilon}
    \end{align}
\end{subequations}
where the gradient of $\mathcal{L}^{\text{MSE}}$ is straightforward:
\begin{equation}
    \label{def:: prediction gradient}
    \frac{\partial \mathcal{L}^{\text{MSE}}}{\partial \Theta} = \frac{1}{N_c} \sum\nolimits_{i \in \mathcal{I}_{c}} 2(\Theta^{\top} x_i - y_i) x_i.
\end{equation}

The gradients of $\mathcal{L}^{\text{TaskI}}$ w.r.t. $\Theta$ and $\epsilon$ require differentiation through an optimization model. Following the chain rule and using $G^{*}$ as an intermediate variable, we obtain:
\begin{subequations}
    \label{def:: day ahead gradient}
    \begin{align}
        \frac{\partial \mathcal{L}^{\text{TaskI}}}{\partial \Theta} &= \frac{1}{N_c} \sum\limits_{i \in \mathcal{I}_{c}}\frac{\partial \mathcal{L}^{\text{TaskI}}_i}{\partial G_i^{*}}\frac{\partial G_i^{*}}{\partial \Theta} =\frac{1}{N_c} \sum\limits_{i \in \mathcal{I}_{c}} C^{\top} \frac{\partial G_i^{*}}{\partial \Theta}\label{def::day ahead task gradient theta} \\
        \frac{\partial \mathcal{L}^{\text{TaskI}}}{\partial \epsilon} &= \frac{1}{N_c} \sum\limits_{i \in \mathcal{I}_{c}}\frac{\partial \mathcal{L}^{\text{TaskI}}_i}{\partial G_i^{*}}\frac{\partial G_i^{*}}{\partial \epsilon} = \frac{1}{N_c} \sum\limits_{i \in \mathcal{I}_{c}} C^{\top} \frac{\partial G_i^{*}}{\partial \epsilon}\label{def::day ahead task gradient epsilon}
    \end{align}
\end{subequations}
where $C = stack(c_g, c_r, c_{r})$ aggregates the cost parameters associated with $g$, $r^{\Plus}$, and $r^{\Minus}$. For the second part of the gradient,  ${\partial G_i^{*}}/{\partial \epsilon}$ can be directly computed using the differentiable optimization layers proposed in \cite{amos2017optnet, agrawal2019differentiable}, while ${\partial G_i^{*}}/{\partial \Theta}$ requires further decomposition, as:
\begin{equation}
    \label{def:: layer}
    \frac{\partial G_i^{*}}{\partial \Theta} = \frac{\partial G_i^{*}}{\partial \hat y_i}\frac{\partial \hat y_i}{\partial \Theta} = \frac{\partial G_i^{*}}{\partial \hat y_i} x_i, 
\end{equation}
where ${\partial G_i^*}/{\partial \hat y_i}$ can be directly computed by the layers.

Similarly, the gradients of $\mathcal{L}^{\text{TaskII}}$ can be written as:
\begin{subequations}
    \label{def:: real time gradient}
    \begin{align}
        \frac{\partial \mathcal{L}^{\text{TaskII}}}{\partial \Theta} &= \frac{1}{N_c} \sum\nolimits_{i \in \mathcal{I}_{c}} \frac{\partial \mathcal{L}^{\text{TaskII}}_i}{\partial G_i^{*}} \frac{\partial G_i^{*}}{\partial \Theta} \label{def:: realtime task gradient theta} \\
        \frac{\partial \mathcal{L}^{\text{TaskII}}}{\partial \epsilon} &= \frac{1}{N_c} \sum\nolimits_{i \in \mathcal{I}_{c}} \frac{\partial \mathcal{L}^{\text{TaskII}}_i}{\partial G_i^{*}} \frac{\partial G_i^{*}}{\partial \epsilon} \label{def:: realtime task gradient epsilon}
    \end{align}
\end{subequations}
We apply the Envelope Theorem \cite{takayama1985mathematical} (detailed in Appendix~\ref{appendix:Envelop Theorem}) to reformulate the first part of these gradients. This reformulation uses $D^*$, the optimal dual solution of \eqref{RT:: total}, as:
\begin{subequations}
\allowdisplaybreaks
    \label{def:: real time gradient calculation}
    \begin{align}
        \frac{\partial \mathcal{L}^{\text{TaskII}}_i}{\partial g^{*}_i} &= \lambda_i^{*}\mathbf{1} - \underline{\mu}_i^{*} + \overline{\mu}_i^{*} + S^{\top}\Phi^{\top}(\overline{\phi}_i^* - \underline{\phi}_i^*) \label{def:: realtime gradient of generators} \\
        \frac{\partial \mathcal{L}^{\text{TaskII}}_i}{\partial r^{\Plus*}_i} &= - \overline{\nu}_i^* \label{def:: realtime gradient of upper reserve} \\
        \frac{\partial \mathcal{L}^{\text{TaskII}}_i}{\partial r^{\Minus*}_i} &= - \underline{\nu}_i^* \label{def:: realtime gradient of lower reserve}
    \end{align}
\end{subequations}
We summarize the proposed end-to-end forecast model calibration approach in Algorithm~\ref{alg:: DR end-to-end algorithm}.
\begin{algorithm}[t]
    \caption{End-to-End Forecast Model Calibration}
    \label{alg:: DR end-to-end algorithm}
    \SetAlgoLined
    \SetKwInOut{Input}{input}\SetKwInOut{Output}{output}
    \Input{Uncertainty quantification $\hat{\mathbb{P}}_{j}$, $j=1,...N_w$;
    Calibration dataset $\mathcal{D}_{c}$; Stop criteria $\Delta \mathcal{L}_{\min}$;
    Learning rates $\kappa_{\theta}$ and $\kappa_{\epsilon}$.\\
    }
    \Output {Calibrated forecast model parameter $\Theta$; \\
    Prescribed decision conservativenss $\epsilon$.
    }
    \Begin{ 
      Initialization: $\Theta \leftarrow \Theta_0$, $\epsilon \leftarrow \epsilon_0$;\\
      Construct $\mathcal{A}^{\text{MSW}}$ with $\hat{\mathbb{P}}_{j}, \forall j$ and $\epsilon$ following \eqref{def:: MSW Ambiguity set};\\
       
      \While{$|\Delta \mathcal{L}| \geq \Delta \mathcal{L}_{\min}$}
      { \For{$(x_i, y_i) \in \mathcal{D}_{c}$}
        {     
            Calculate forecast $\hat y_i = \Theta^\top x_i$;\\
            Solve \eqref{RDA:: total} with $\hat y_i$ and $\mathcal{A}^{\text{MSW}}$ for $G_i^{*}$; \\
            Derive $\frac{\partial G_i^{*}}{\partial \Theta}$ and $\frac{\partial G_i^{*}}{\partial \epsilon}$ with cvxpylayers;\\
            Solve \eqref{RT:: total} with $y_i$ and $G_i^{*}$ for $R_{i}^{*}$ and $D_i^{*}$;\\
        }
        Calculate gradients following \eqref{def:: total loss gradient}-\eqref{def:: real time gradient calculation};\\
        $\Delta \Theta = \nabla_\Theta \mathcal{L}(\Theta, \epsilon)$, 
        $\Delta \epsilon = \nabla_\epsilon \mathcal{L}(\Theta, \epsilon)$;\\
        $\Theta \leftarrow \Theta - \kappa_{\theta} \Delta \Theta$,
        $\epsilon \leftarrow \epsilon_{i} - \kappa_{\epsilon} \Delta \epsilon$.
      }
      \KwRet{$\Theta$, $\epsilon$}
     }
\end{algorithm}

\subsection{Distributed Forecast Model Calibration} 
\label{sec:: Distributed Model Calibration}
\begin{figure}[!htbp]
    \centering
    \includegraphics[width=1\linewidth]{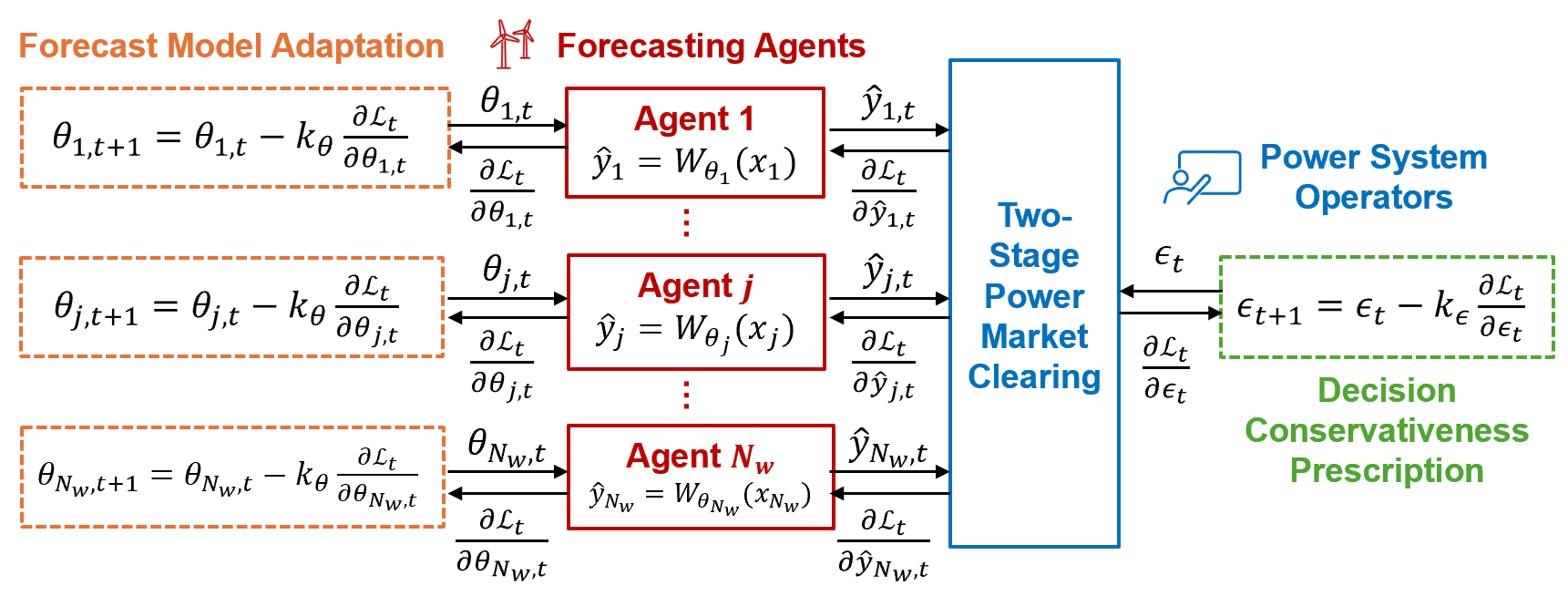}
    \caption{Forecast Model Calibration in a Distributed Manner.}
    \label{fig:distributed_method}
\end{figure}

Algorithm~\ref{alg:: DR end-to-end algorithm} assumes that both forecast model adaptation and decision conservativeness prescription are carried out by system operators. However, in practice, system operators may only have access to the forecasted power $\hat y$, while the forecast model parameters $\Theta$ remain private to wind forecasting agents. In such cases, forecast model calibration can be performed in a distributed manner by the forecasting agents, as illustrated in Fig.~\ref{fig:distributed_method}. System operators must provide each forecasting agent $j$ with the calibration dataset $\mathcal{D}_{c}$ and use the forecasted wind power $\hat y_{j,i}, i \in \mathcal{I}_{c}$ to clear the two-stage market in the forward pass. In the backward pass, system operators need to send $\partial \mathcal{L}/ \partial \hat y_j$ to each agent $j$, allowing the agent compute the second part of the gradient, i.e., $\partial \hat y_j/ \partial \theta_j$. This approach accommodates the use of different forecast models by different agents, while also reducing the computational burden on the system operators.

\section{Numerical Experiments}
This section presents numerical experiment results to demonstrate the effectiveness of the proposed method. The tests are conducted on the IEEE 5-bus system, adapted from Pandapower \cite{pandapower2018}. One wind farm is added to node 3 with a maximum capacity of 200 MW. Two features are considered in the wind power forecast model, i.e., $\Theta$ has a dimension of 2. All numerical experiments are performed in Python using the CVX solver \cite{cvx} on a PC equipped with an Intel Core i9 processor (2.20 GHz) and 16 GB of RAM. The cvxpylayer package \cite{agrawal2019differentiable} is used to compute the gradients of $G^*$ with respect to $\Theta$ and $\hat y$. 

\begin{table}[!t]
    \centering
    \renewcommand{\arraystretch}{1.2}
    \caption{Parameter Setting in the Numerical Experiment}
    \begin{tabular}[width=0.8\textwidth]{cc|cc}
    \hline
    \textbf{Parameter} & \textbf{Value} & \textbf{Parameter} & \textbf{Value} \\ \hline
    $N_s$              & 20             & $c_g$             & [14, 30, 10] \$/MW   \\ 
    $N_c$              & 20             & $c_r$             & $1.5 \cdot c_g$ \\ 
    $\kappa_{\theta}$  & $1.00 \times 10^{-4}$ & $c_a = c_{\text{in}}$ & $2 \cdot c_g$   \\ 
    $\kappa_{\epsilon}$ & $1.00 \times 10^{-3}$ & $c_{\text{out}}^{\Plus} = c_{\text{out}}^{\Minus}$ & $10 \cdot c_g$ \\ 
    $\epsilon_0$       & 1              & $\gamma$  & 0.05  \\ 
    $\Theta_0$         & [1, 2]         & $\overline{\xi} = -\underline{\xi}$          & 50 MW           \\ \hline
    \end{tabular}
    \label{Table:parameters}
\end{table}

\begin{table}[!t]
    \centering
    \renewcommand{\arraystretch}{1.2}
    \caption{Optimal Values of $\Theta$ and $\epsilon$ under different $\sigma_c$}
    \begin{tabular}[width=0.8\textwidth]{c|ccccc}
    \hline
    $\sigma_c$      & 15 & 18 & 20  & 22 & 25 \\ \hline
    $\epsilon^*$    & 0.176 & 0.225 & 1.102 & 1.175 & 1.427 \\ 
    $\Theta^*[1]$   & 0.962 & 0.944 & 0.925 & 0.913 & 0.901 \\ 
    $\Theta^*[2]$   & 2.085 & 2.113 & 2.158 & 2.187 & 2.236 \\ \hline
    \end{tabular}
    \label{Table:optimal value}
\end{table}

The parameter settings for the numerical experiment are summarized in Table~\ref{Table:parameters}. Uniformly distributed $N_s$ and $N_c$ samples of $x$ are generated for the UQ and calibration datasets, respectively. The corresponding $\Theta_0^\top x$ values are computed, and the actual $y$ values are obtained by adding a normally distributed prediction error $\hat{\xi}$. In the UQ dataset, $\hat{\xi} \sim \mathcal{N}(0, \sigma_s^2)$ with $\sigma_s = 10$, while in the calibration dataset, $\hat{\xi} \sim \mathcal{N}(0, \sigma_c^2)$, with $\sigma_c$ values listed in Table~\ref{Table:optimal value}.

Table~\ref{Table:optimal value} shows the optimal values of $\Theta$ (for both dimensions) and $\epsilon$ obtained using the proposed framework under varying values of $\sigma_c$. In this experiment, the weight parameter $\eta = 1$, balancing task loss and forecast error. As $\sigma_c$ increases, the gap between the actual distribution of the forecast error $\xi$ and the empirical distribution widens. This necessitates greater decision conservativeness in look-ahead scheduling, represented by larger $\epsilon$ values. This trend is consistent with the results for $\epsilon^*$ in Table~\ref{Table:optimal value}.

Furthermore, larger values of $\sigma_c$ indicate a need for increased system flexibility, resulting in higher task losses. With a fixed weight $\eta$, the optimization process places greater emphasis on minimizing task loss. Consequently, the calibrated parameter $\Theta^*$ deviates more significantly from the initial $\Theta_0$, which represents the optimal $\Theta$ for minimizing forecast error alone. This deviation is reflected in Table~\ref{Table:optimal value}, where the difference between $\Theta^*$ and $\Theta_0$ grows with increasing $\sigma_c$.

\begin{figure}[!htbp]
    \centering
    \includegraphics[width=1\linewidth]{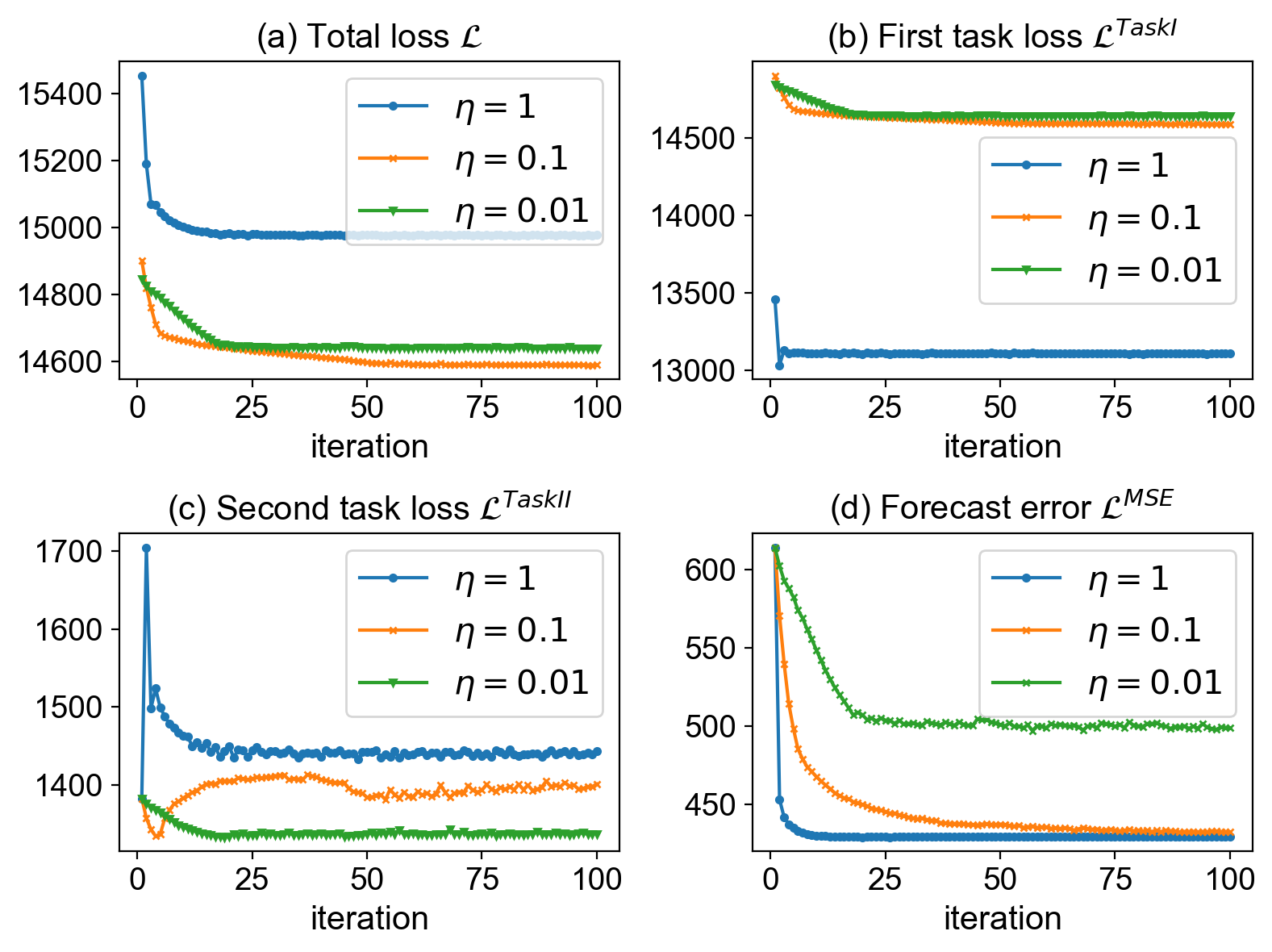}
    \caption{Convergence Analysis of Loss Functions for Different Weights $\eta$}
    \label{fig:converge}
\end{figure}

Fig.~\ref{fig:converge} illustrates the convergence behavior of $\mathcal{L}$, $\mathcal{L}^{\text{TaskI}}$, $\mathcal{L}^{\text{TaskII}}$, and $\mathcal{L}^{\text{MSE}}$ across iterations for different values of the weighting parameter $\eta$. Larger values of $\eta$ prioritize minimizing the forecast error, leading to faster convergence across all loss types. Conversely, smaller values of $\eta$ result in slower and less significant reductions in forecast error but achieve lower task losses. This demonstrates the trade-off between task loss and forecast error and highlights the influence of $\eta$ on the optimization dynamics and final convergence outcomes.

\section{Conclusion and Future Work}
\label{sec:conclusion}
This paper presents an end-to-end framework for power system operators to calibrate wind power forecast model parameters and determine the optimal decision conservativeness while minimizing downstream costs. This task is efficiently achieved using gradient descent and can be implemented in a distributed manner by forecasting agents. Numerical experiments on the IEEE 5-bus system demonstrate the effectiveness of the proposed framework, along with a sensitivity analysis of the results to parameter settings, such as the weight between task loss and forecast error. The framework is versatile and can be easily extended to other forecasting tasks in power systems. Future research will investigate the performance of this end-to-end framework in larger systems with more forecasting agents and more complex forecast models. 
\bibliographystyle{IEEEtran}
\bibliography{ref}

\appendix
\subsection{Envelop Theorem}
\label{appendix:Envelop Theorem}
\begin{myth} \label{Thm:: Envelop Theorem}
    Consider the following parametric optimization problem, where $f(x, \alpha)$ and $g_j(x, \alpha), j=1,...,m$ are real-valued functions that are continuously differentiable with respect to both $x$ and $\alpha$:
    \begin{subequations}
        \label{def:: Envelop Theorem opt}
        \begin{align}
            \max_{x}\ \qquad  & f(x, \alpha) \\
            \mathrm{s.t.} \ (\lambda_j) \ & g_j(x,\alpha) \geq 0, j=1,\dots,m
        \end{align}
    \end{subequations}
    Let $x^*(\alpha)$ and $\lambda_j(\alpha)$ be the optimal primal and dual solutions of \eqref{def:: Envelop Theorem opt} and define the value function as $V(\alpha) = f(x^{*}(\alpha), \alpha)$. The gradient of the value function $V(\alpha)$ with respect to $\alpha$ is given by:
    \begin{equation}
        \label{def:: Envelop Theorem gradient}
        \frac{\partial V(\alpha)}{\partial \alpha} = \frac{\partial f (x^*(\alpha), \alpha)}{\partial \alpha} + \sum\nolimits_{j=1}^{m} \lambda_{j}^{*}(\alpha) \frac{\partial g_j(x^*(\alpha), \alpha)}{\partial \alpha}
    \end{equation}
\end{myth}
The proof of this theorem can be found in \cite{takayama1985mathematical}.

\end{document}